# Multiphase Astrophysics to Unveil the Virgo Environment (MAUVE)

Barbara Catinella[1]
Luca Cortese[1]
Jiayi Sun[2]
Toby Brown[3]
Eric Emsellem[4]
Amelia Fraser-McKelvie[4]
Adam B. Watts[1]
Amy Attwater[1]
Andrew Battisti[1,5]
Alessandro Boselli[6]
Woorak Choi[7]
Aeree Chung[8]
Elisabete da Cunha[1]
Timothy A. Davis[9]
Sara Ellison[10]
Pavel Jáchym[11]
Maria J. Jimenez-Donaire[12,13]
Tutku Kolcu[14]
Bumhyun Lee[8]
James McGregor[5]
Ian Roberts[15]
Eva Schinnerer[16]
Kristine Spekkens[17]
Sabine Thater[18]
David Thilker[19]
Jesse van de Sande[20]
Vicente Villanueva[21]
Thomas G. Williams[22]
Nikki Zabel[23]

[1] International Centre for Radio Astronomy Research, The University of Western Australia, Australia
[2] Department of Astrophysical Sciences, Princeton University, USA
[3] National Research Council of Canada, Herzberg Astronomy and Astrophysics Research Centre, Canada
[4] ESO
[5] Research School of Astronomy and Astrophysics, Australian National University, Australia
[6] Aix Marseille University, CNRS, CNES, LAM, France
[7] Department of Physics and Astronomy, McMaster University, Canada
[8] Department of Astronomy, Yonsei University, Republic of Korea
[9] Cardiff Hub for Astrophysics Research & Technology, School of Physics & Astronomy, Cardiff University, UK
[10] Department of Physics & Astronomy, University of Victoria, Canada
[11] Astronomical Institute of the Czech Academy of Sciences, Czech Republic
[12] AURA for ESA, Space Telescope Science Institute, USA
[13] National Astronomical Observatory (IGN), Madrid, Spain
[14] School of Physics and Astronomy, University of Nottingham, UK
[15] Department of Physics & Astronomy and Waterloo Centre for Astrophysics, University of Waterloo, Canada
[16] Max Planck Institute for Astronomy, Heidelberg, Germany
[17] Department of Physics, Engineering Physics and Astronomy, Queen's University, Canada
[18] Department of Astrophysics, University of Vienna, Austria
[19] Department of Physics and Astronomy, Johns Hopkins University, Baltimore, USA
[20] School of Physics, University of New South Wales, Australia
[21] Department of Astronomy, Concepción University, Chile
[22] Sub-department of Astrophysics, Department of Physics, University of Oxford, UK
[23] Department of Astronomy, University of Cape Town, South Africa

The Multiphase Astrophysics to Unveil the Virgo Environment (MAUVE) project is a multi-facility programme exploring how dense environments transform galaxies. Combining a VLT/MUSE P110 Large Programme and ALMA observations of 40 late-type Virgo Cluster galaxies, MAUVE resolves star formation, kinematics, and chemical enrichment within their molecular gas discs. A key goal is to track the evolution of cold gas that survives in the inner regions of satellites after entering the cluster, and how it evolves across different infall stages. With its high spatial resolution — probing down to the physical scales of giant molecular cloud complexes — and multiphase synergy, MAUVE aims to offer a time-resolved view of environmental quenching and set a new benchmark for cluster galaxy studies.

## The puzzle of environmental quenching

Galaxy clusters are among the most extreme and transformative environments in the Universe. As galaxies fall into these dense regions, they experience a variety of physical processes — including ram pressure stripping, tidal interactions, starvation and feedback — that can rapidly reshape their morphology, gas content and star formation activity. While the broad strokes of environmental quenching are well established, particularly the outside-in removal of gas (for example, Cayatte et al., 1990; Chung et al., 2009; Cortese et al., 2021), much remains unknown about how star formation ceases in the inner discs of satellite galaxies, which often retain significant reservoirs of cold gas long after the interstellar medium in their outer discs has been stripped.

The Multiphase Astrophysics to Unveil the Virgo Environment (MAUVE)[1] project addresses this challenge by focusing on the inner discs of late-type galaxies that are currently being transformed by their environment. By targeting a mass-limited sample of Virgo Cluster galaxies across the full infall sequence, MAUVE is designed to capture the critical phase where stripping, feedback, and internal dynamics begin to alter the cold gas–star formation cycle from within.

## The MAUVE survey: MUSE and ALMA in synergy

At the heart of MAUVE is the coordinated use of the Very Large Telescope Multi Unit Spectroscopic Explorer (MUSE) and the Atacama Large Millimeter/submillimeter Array (ALMA) to link stellar populations, ionised gas and cold molecular gas within the same galaxies. This synergy enables a systematic, spatially-resolved investigation of how cluster environments influence both the fuelling and suppression of star formation across the entire disc — not just in the outer regions of galaxies where stripping is most obvious, but also in their dense, star-forming centres.

The MAUVE sample comprises 40 late-type galaxies in the Virgo Cluster spanning a broad range of infall stages, from first-infallers to post-pericentre systems. The sample is rooted in the VIVA survey (Chung et al., 2009), which mapped the atomic hydrogen (H I) gas content of 53 Virgo spirals using the Very Large Array. These data provide critical insight into gas removal and stripping signatures in the outer discs of satellites.

Building on VIVA, the VERTICO[2] survey (Brown et al., 2021) used the ALMA Atacama Compact Array (ACA) to obtain

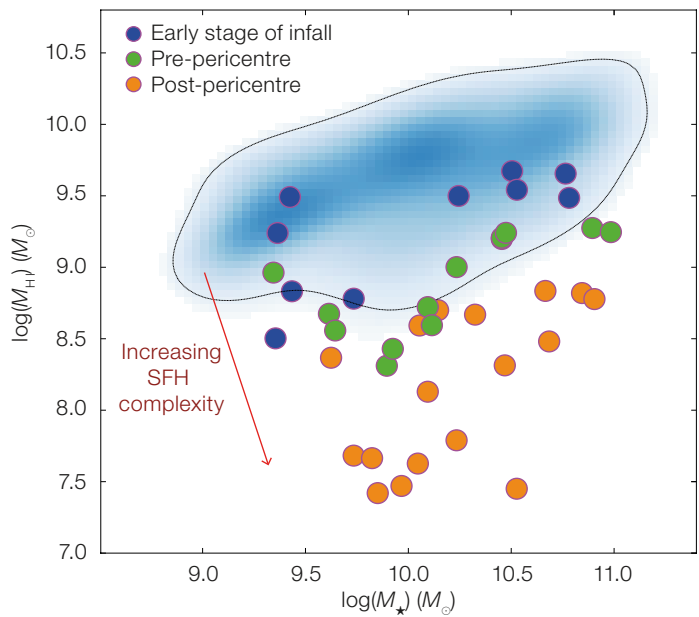

Figure 1. The MAUVE sample, spanning the full infall sequence of Virgo Cluster satellite galaxies. H I mass is plotted against stellar mass, with galaxies colour-coded by infall stage: early infallers (blue), approaching first pericentre (green), and post-pericentre systems (orange). The blue shaded region shows local star-forming galaxies from the extended GALEX Arecibo SDSS Survey (xGASS; Catinella et al., 2018).

Figure 1 illustrates the sample selection in the H I–stellar mass plane, colour-coded by infall stage. Early infallers show modest H I depletion relative to isolated galaxies, while post-pericentre systems have lost most of their atomic gas but typically retain significant molecular reservoirs. These surviving cold gas discs offer a unique opportunity to investigate where, when and how star formation is ultimately shut down — a key reason why they comprise the largest population in the sample.

The trend of gas depletion across infall stage forms the backbone of the MAUVE strategy: to trace how galaxies evolve not just globally, but also internally as they traverse the cluster potential. In short, MAUVE is designed to systematically map representative paths of environmental transformation — from fuelling to quenching, from gas to stars — at the scales where these processes physically unfold.

CO(2–1) maps for 51 of the 53 VIVA galaxies at spatial resolutions of 600–800 pc, tracing the cold, star-forming molecular gas. MAUVE targets all VERTICO galaxies with stellar mass $M_\star > 10^{9.3}\,M_\odot$ and a detection in CO(2–1), resulting in a mass-limited, molecular-gas-selected sample well suited to resolved environmental studies.

The ALMA data from VERTICO are complemented by MAUVE's new ALMA Band 6 programme, which combines 12-metre-array and ACA observations to deliver imaging of CO(2–1) emission at a resolution of around 50 pc. These data enable exquisitely detailed studies of molecular gas structure, dynamics and gas depletion times across the disc.

MUSE observations complement the millimetre data by mapping each galaxy's stellar continuum and ionised gas emission — along with their respective kinematics — at around 100–200-pc resolution. All galaxies are observed to sufficient depth to support spectral fitting and emission-line diagnostics, with one to nine pointings per target (four on average), for a total of 145 pointings and 173 hours of observing time. Pointings are designed to fully cover the molecular gas disc, ensuring optimal spatial overlap with the ALMA data.

Taken together, these observations provide a uniquely deep and spatially resolved, multiphase view of galaxy evolution in a massive cluster — akin to what the Physics at High Angular resolution in Nearby Galaxies (PHANGS) project (Leroy et al., 2021; Emsellem et al., 2022) has achieved for field galaxies.

### Building a time sequence of environmental transformation

A core strength of MAUVE lies in its sample design, which captures not just the diversity of late-type galaxies in clusters, but the process of transformation itself. Rather than selecting galaxies based on specific features or extreme morphologies, MAUVE targets a representative range of systems that collectively span the accretion histories of galaxies infalling into the cluster.

Crucially, for each of the 40 MAUVE galaxies the stage of infall — whether the galaxy is a recent arrival, approaching first pericentre, or has already passed it — has been determined using a combination of orbital modelling (for example, Vollmer, 2009) and gas morphology diagnostics (Yoon et al., 2017). This enables MAUVE to move beyond snapshot studies and to assemble a time sequence of environmental influence, linking physical processes to evolutionary stages. Inevitably, the significant improvement in data quality will also allow a revision of previously estimated infalling stages and it is possible that we will be finding some surprises.

### What drives quenching in Virgo?

Through its spatially resolved, multiphase view of Virgo Cluster galaxies, MAUVE aims to advance our understanding of several key processes that regulate star formation and structural evolution in dense environments.

### Where, and how, does star formation shut down?

The sustained star formation in gas-rich inner discs of satellite galaxies poses a key challenge to models of quenching. MAUVE addresses this by combining H$\alpha$ emission — sensitive to short-timescale star formation (around 10 Myr) — with full spectral fitting of the stellar continuum to reconstruct star formation histories (SFH) over the past several Gyr. This approach reveals how quenching progresses spatially and temporally, especially within molecular gas discs that survive initial stripping.

### How is the molecular gas reservoir affected during galaxy infall?

Molecular gas is the direct fuel for star formation, yet its response to environmental effects remains poorly constrained. ALMA data allow MAUVE to trace the structure, surface density and dynamics of the molecular disc at around 50-pc resolution, resolving giant molecular cloud complexes. These data can be used to test whether stripping penetrates into molecular-rich regions, whether gas becomes compressed or turbulent, and whether the efficiency of converting $H_2$ to stars changes during galaxy infall. These observations, combined with MUSE-based star formation tracers, provide direct measurements of local star formation efficiency and the full gas-star formation-feedback cycle across the disc and under environmental pressure.

Do feedback and outflows contribute to quenching?

Feedback from massive stars and supernovae is a critical regulator of star formation in field galaxies, but its role in cluster environments is less well understood. MAUVE has already identified ionised gas outflows in several galaxies, including systems with surprisingly low star formation rates. A particularly striking case is shown in Figure 2: MUSE observations of NGC 4383 reveal a roughly 6-kpc bipolar outflow emerging from the central regions of the galaxy (Watts et al., 2024), providing strong evidence that feedback can remain effective in the dense environment of a cluster. Whether such outflows are common, and how they vary with infall stage, are questions that MAUVE is uniquely positioned to address.

How does the environment reshape galaxy dynamics and morphology?

Beyond removing gas and quenching star formation, cluster environments can affect the internal dynamics of galaxies — thickening discs, altering stellar orbits, or driving morphological transitions toward earlier types. These dynamical effects often unfold over longer timescales than gas stripping and may only become apparent after a galaxy has passed pericentre. MAUVE combines MUSE maps of stellar and ionised gas kinematics with ALMA molecular gas data to capture the full dynamical response of galaxies to their environment. These observations can reveal disturbances such as warped velocity fields, asymmetric rotation, enhanced dispersion, or misaligned components — signatures of tidal interactions, disc heating, or structural reconfiguration during infall. The combination of these kinematic tracers and resolved stellar populations will allow MAUVE to link dynamical evolution with SFHs and environmental exposure, offering a comprehensive view of how galaxies are physically transformed in clusters.

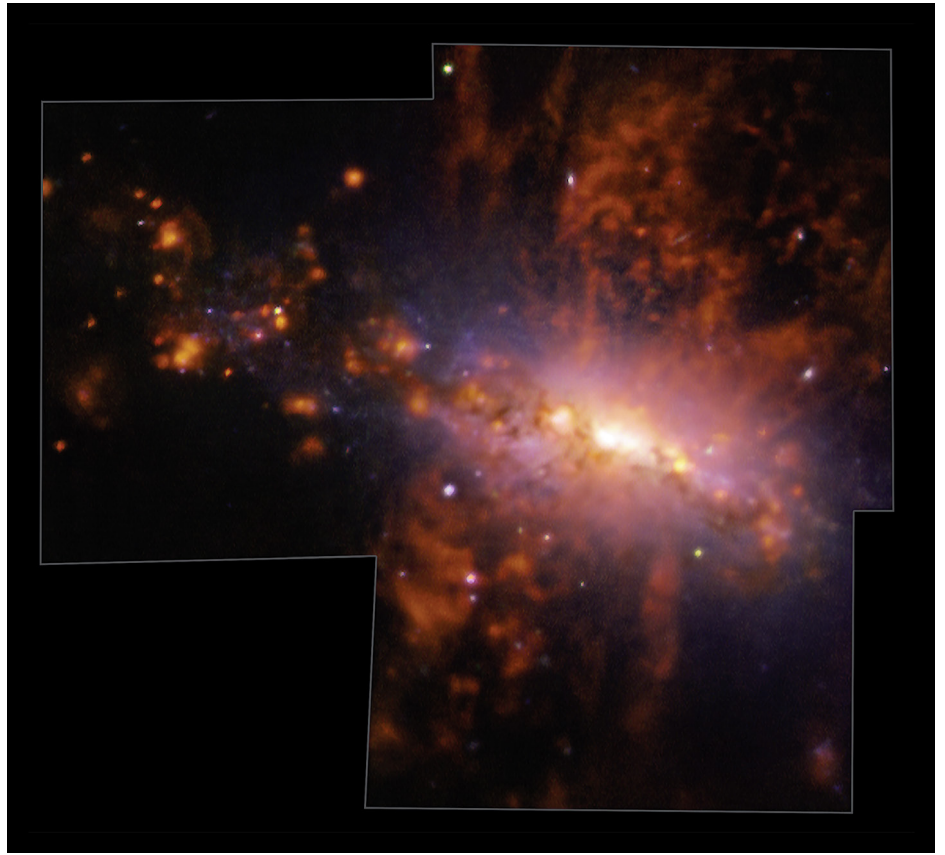

Figure 2. MAUVE–MUSE observations of NGC 4383 showing a spectacular outflow of ionised gas (Hα, red), superposed on a colour image of the galaxy (Watts et al., 2024). The image covers an area of 8.8 × 8 kpc at a distance of 16.5 Mpc.

High-resolution mapping of environmental impact

Two of MAUVE's defining strengths are the high resolution and uniformity of its multiphase observations. As illustrated in Figure 3, the combination of MUSE and ALMA delivers sharp, high-signal-to-noise maps of ionised and molecular gas, revealing the internal structure and kinematics across the full extent of the star-forming disc in exquisite detail. The ability to jointly interpret these components — for example, by comparing ionised and molecular gas kinematics, or tracing metallicity gradients in relation to CO morphology — is essential for identifying where and how environmental processes are at work.

Observations for the MUSE P110 Large Programme began in early 2023 and have progressed steadily. At the time of writing, 30 of the 40 galaxies have been observed (and an additional three have PHANGS archival data), bringing the programme to 76% completion in terms of observing time. If the current pace continues, MAUVE–MUSE should be completed by mid-2026. The new MAUVE–ALMA observations have been conducted during Cycles 10 and 11 and, including archival data, a total of 33 galaxies have been completed to date.

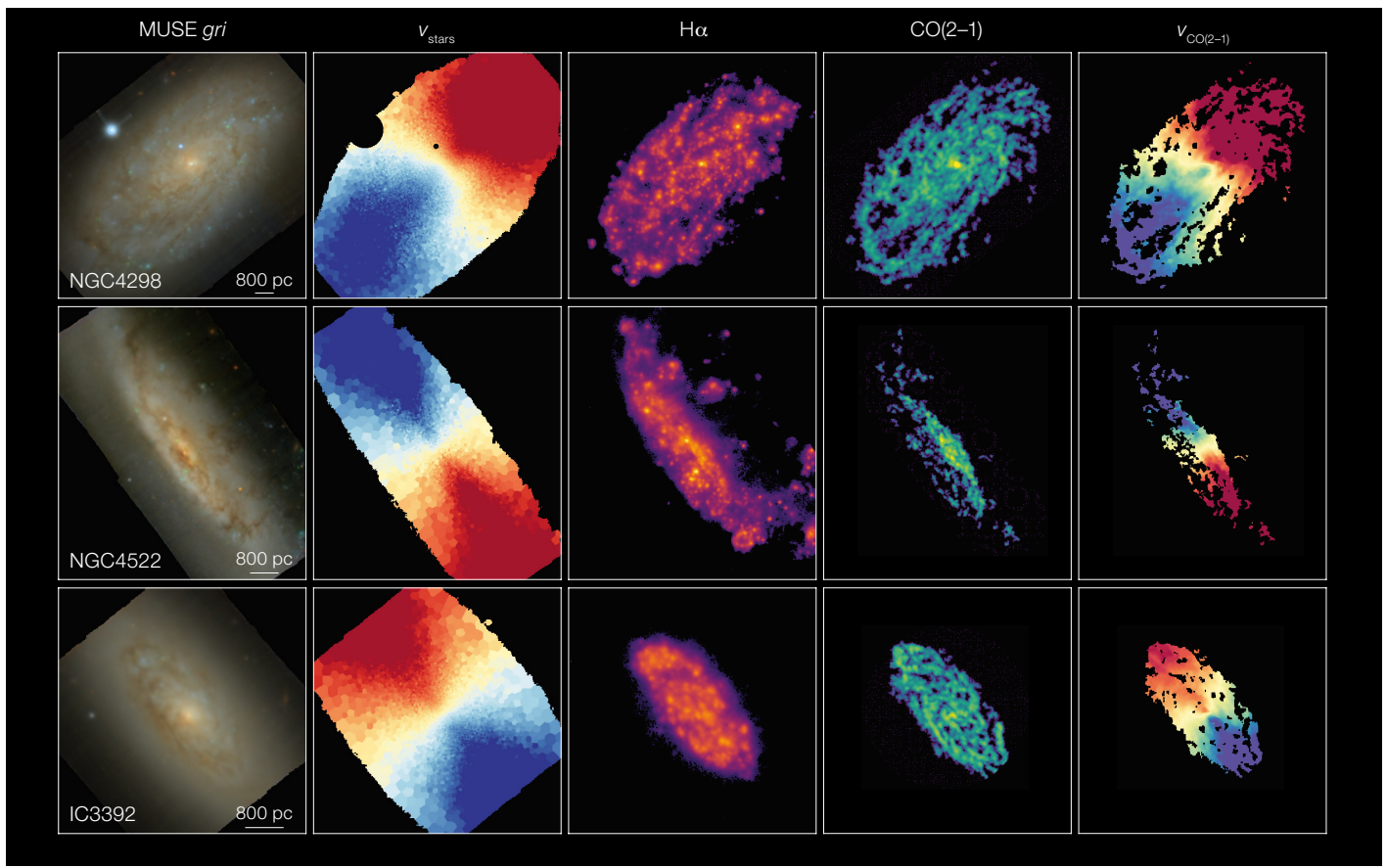

Figure 3. Montage showing MAUVE data quality for three representative galaxies at different stages of cluster infall, from early infall (top) to post-pericentre (bottom). For each galaxy, the panels (left to right) show the MUSE *gri* composite image, stellar velocity field, Hα intensity map, and ALMA CO(2–1) integrated intensity and velocity maps. The horizontal white bar indicates 10 arcseconds, corresponding to about 800 pc at Virgo's distance of 16.5 Mpc. Together, the MUSE and ALMA data reveal the internal structure and kinematics of stars and gas across the molecular disc at high spatial resolution.

## Summary and outlook

MAUVE provides the highest spatial resolution to date in a multiphase study of star-forming galaxies across the Virgo Cluster's infall sequence. By combining optical and millimetre observations of gas, stars, feedback and kinematics within the same systems, it offers an unprecedented view of how environmental processes act on scales as small as 50 pc, of the order of giant molecular cloud complexes, and as a function of infall time. These exquisite data will allow us to map where and when star formation is quenched, how molecular gas is displaced or stripped and what role feedback and outflows play in regulating or accelerating this transformation.

Its findings will inform both the interpretation of high-redshift surveys and the development of state-of-the-art hydrodynamical simulations, many of which still struggle to reproduce the detailed interplay between environment and interstellar medium. As the programme progresses, MAUVE is expanding its scope through coordinated multi-wavelength efforts. Notably, a new Hubble Space Telescope Cycle 33 Treasury programme (MAUVE–HST) was recently approved. It will deliver deep UV-to-optical imaging for all 40 galaxies in the sample, enabling detailed characterisation of stellar clusters, associations, luminous field stars, and H II regions. With these new data and future synergies across ground- and space-based facilities, MAUVE is poised to become a foundational dataset for understanding how dense environments transform galaxies from within.

#### Acknowledgements

We would like to thank the ESO and ALMA staff for their substantial and ongoing support throughout the MAUVE observing programme.

#### References

Brown, T. et al. 2021, ApJS, 257, 21
Catinella, B. et al. 2018, MNRAS, 476, 875
Cayatte, V. et al. 1990, AJ, 100, 604
Chung, A. et al. 2009, AJ, 138, 1741
Cortese, L. et al. 2021, PASA, 38, 35
Emsellem, E. et al. 2022, A&A, 659, A191
Leroy, A. K. et al. 2021, ApJS, 257, 43
Vollmer, B. 2009, A&A, 502, 427
Watts, A. B. et al. 2024, MNRAS, 530, 1968
Yoon, H. et al. 2017, ApJ, 838, 81

#### Links

[1] MAUVE: https://mauve.icrar.org
[2] VERTICO: https://sites.google.com/view/verticosurvey